# Structural, Thermal and Electrical properties of Poly(methyl methacrylate)/$CaCu_3Ti_4O_{12}$ composite sheets fabricated via melt mixing.

P. Thomas,[a]* R.S. Ernest Ravindran.[a]  K.B.R. Varma.[b]

[a] Dielectric Materials Division, Central Power Research Institute, Bangalore : 560 080, India

[b] Materials Research Centre, Indian Institute of Science, Bangalore: 560012, India

**Abstract**

Poly(Methyl Methacrylate) (PMMA) and $CaCu_3Ti_4O_{12}$ (CCTO) composites were fabricated via melt mixing followed by hot pressing technique. These were characterized using X-ray diffraction (XRD), thermo gravimetric (TGA), Thermo Mechanical (TMA), Differential scanning calorimetry (DSC), Fourier transform infrared (FTIR), and Impedance analyser for their structural, thermal and dielectric properties. Composites were found to have better thermal stability than that of pure PMMA. However, there was no significant difference in the glass transition $(T_g)$ temperature between the polymer and the composite. The appearance of additional vibrational frequencies in the range 400-600 $cm^{-1}$ in FTIR spectra indicated a possible interaction between PMMA and CCTO. The composite, with 38 Vol % of CCTO (in PMMA), exhibited remarkably low dielectric loss at high frequencies and the low frequency relaxation is attributed to the space charge polarization/MWS effect. The origin of AC conductivity particularly in the high frequency region was attributed to the electronic polarization.

---

* Corresponding author : Tel. +91-80-2360 1454; Fax: +91-80-2360-4448.
E-mail : thomas@cpri.in (P.Thomas)

## INTRODUCTION

Recently, much research work has been done on polymer-ceramic composites keeping potential industrial applications in vies.The hybrid materials comprising of ceramic fillers and polymer matrix have drawn widespread attention of researchers. Improvement in the mechanical and thermal properties have been achieved by introducing fillers into flexible polymer matrices [1,2]. Owing to the continuous development towards the miniaturization of electronics, high dielectric constant polymer-ceramic composites have become increasingly promising materials for embedded capacitor applications [3-7]. These embedded capacitors are part of printed circuit board (PCB) laminates and fabrication of such devices is a major challenge. These embedded capacitors apart from exhibiting high capacitance, require to possess low thermal coefficient of permittivity (TCK) over a wide range of temperatures. In this regard, $CaCu_3Ti_4O_{12}$ (CCTO) ceramic whose dielectric constant is nearly independent of frequency (upto 10 MHz) accompanied by low thermal coefficient of permittivity (TCK) in the 100-600K temperature range [8,9] has been used as a filler and studied [10-17] to explore the possibility of obtaining a new generation composites associated with high permittivity for capacitor applications in electrical circuits. The integration of embedded capacitors in the inner layer of PCBs also needed to be processed at low temperatures [18]. Hence, it is essential to understand the thermal behaviour of the polymer-ceramic composites as the dielectric behaviour of these composites not only dependent on the type of polymer and its structure, and also dependent on the type of filler, the interface bonding between the polymer and the ceramic, the thermal properties and the heat capacities of the individual components [19,20]. The Poly (Methyl Methacrylate) (PMMA), a transparent thermoplastic polymer, possesses moderate physical properties associated with low cost and the composite systems based on this were studied in great detail [21-34].

The aim of the present work was to fabricate and characterize PMMA+CCTO composites and study their structural, thermal and electrical behaviour using various analytical techniques such as X-Ray Diffraction (XRD), Thermo Gravimetric Analysis (TGA), Differential Scanning Calorimetry (DSC), FTIR spectroscopy, Thermo Mechanical Analysis (TMA) and Impedance analyser.

**EXPERIMENTAL**

*Synthesis of $CaCu_3Ti_4O_{12}$ powders*

The solid-state reaction route was adopted for synthesizing CCTO ceramic powders [8,15]. The stoichiometric amounts of AR grade $CaCO_3$, CuO and $TiO_2$ was weighed, mixed using acetone and ball milled (300rpm) for 5h. The homogeneous mixture thus obtained was dried in an electric oven for about 1h. This stoichiometric mixture was taken in a re-crystallized alumina crucible and heated at $1000^o$C for 10h to obtain phase pure CCTO [8]. In order to get submicron particles, the CCTO powders were ball milled for about 12h using a planetary mill.

*Fabrication of Poly(methyl methacrylate)/$CaCu_3Ti_4O_{12}$ composite sheets*

PMMA, having Molecular weight of 1,10,000 (Make: LG Corporation) was used as matrix material. For the fabrication of composites, initially, the as received PMMA granules were heated at $210^o$C in Brabender Plasticorder (Model: PLE331) till the PMMA granules were thoroughly melted. To this melt, CCTO powder (0 to 38 % by volume) was slowly added and mixed for 20 min at this temperature. The mixture was taken out from the Plasticorder and hot-pressed at this temperature to obtain a sheet of 100 $mm^2$ with 1.0 mm in thickness. The composite became brittle when the ceramic loading was beyond 38 % by volume. Hence, to obtain flexible composites which could be made into a variety of shapes, the ceramic loading has been restricted to a maximum of 38 Vol %. Fig. 1 shows the flowchart depicting various steps involved in the fabrication of PMMA-CCTO Composites.

*Characterization*

To examine the structure, an XPERT-PRO Diffractometer (Philips, Netherlands) was used. Thermo gravimetric (TGA) analyses were done using the TA Instruments (UK, Model: TGA Q500) and Thermo mechanical analyses (TMA) were done using the TA Instruments (UK, Model: TMA Q400). TGA experiments were conducted in Nitrogen atmosphere at a flow rate of 60ml/min, at a heating rate of 10 deg/min. TMA experiments were conducted in Nitrogen atmosphere at a flow rate of 60ml/min, at a heating rate of 5 deg/min  Infrared spectra were recorded using a FTIR/ATR spectrophotometer, Model : Nicolet 6700, (Make : Thermo Scientific). Differential Scanning Calorimeter, (Make: Mettler Toledo, Model : DSC 821e) was employed at a heating rate of $10^0$C /min under nitrogen atmosphere at a flow rate of 60ml/min, using Aluminium pan). An LCR meter (Model: HP4194A) was used for the capacitance measurements as a function of frequency (100Hz–1MHz) at room temperature.

### 3.0 RESULTS AND DISCUSSION

#### *3.1. X-Ray Diffraction studies:*

The X-ray diffraction patterns were recorded on the CCTO ceramic powders, as received PMMA and on the composites that were fabricated. X-ray diffraction pattern revealed that the as received PMMA is semicrystalline in nature (Fig.2(a)). The XRD data obtained for the as prepared CCTO powders (Fig 2(b)) are compared well with the ICDD data (01-075-1149) shown in Fig 2(c) and with that reported earlier [9]. The X-ray diffraction pattern recorded for the PMMA comprising 6 vol % CCTO revealed its composite nature. In the case of PMMA with 38 vol % CCTO composite (Fig.2(e)), XRD pattern revealed only CCTO peaks as the intensities of these are dominant.

#### *3.2. Thermal studies (TGA/TMA):*

In order to examine the thermal stability, thermal analyses were carried out on PMMA+CCTO composites as well as on pure PMMA for comparison. The TGA and the derivative thermogravimetric (DTGA) data obtained for the pure PMMA and the composites are illustrated in Fig.3. It has been observed that there is a change in the thermal degradation behaviour of PMMA with CCTO. The onset of decomposition temperature has increased as the CCTO filler content increased in PMMA. The onset of decomposition temperature (temperature at 10% weight loss) [21] was found to increase for all the composites under study. The decomposition temperature onset accompanied by 10% weight loss for PMMA+6 vol % CCTO is 358.4$^o$C and for the PMMA+38 vol % CCTO, is 374.1$^o$C, while for pure PMMA, it is 353.5$^o$C. The onset of decomposition temperature is higher by 20$^o$C than that of virgin PMMA. All the samples (PMMA and composites) indicated that there is no weight loss upto 270$^o$C and thereafter, the degradation begins. To find out whether there is more than one step decomposition temperature involved, derivative data from the TGA (DTGA) results were plotted (fig.3) and it has been observed that both the virgin PMMA and the composites undergo only single step degradation. It is seen from the derivative thermogram (Fig.4), that there is an increase in the decomposition temperature as the filler content is increased in the PMMA. Overall, the composites have better thermal stability than that of PMMA. Interestingly, the weight loss obtained from the burnt (950$^o$C/24h) out test on a few composite samples by the gravimetric analysis is in good agreement with the experimental values based on TGA.

The pure PMMA and the composites were also subjected to Thermo Mechanical Analysis (TMA), to understand their thermal expansion behaviour as this is crucial for several device applications. Fig.5 show the Coefficient of Thermal Expansion (CTE) as a function of temperature for pure PMMA and the composites. It has been observed that as the CCTO content increased in PMMA, Coefficient of Thermal Expansion (CTE), measured between 40-100$^o$C, has decreased (as compared to that of pure PMMA) and the density of the composites increased. The decrease in the CTE for the composites as the CCTO increased in the PMMA is the indication of good adhesion between the polymer and the ceramic. The glass transition (Tg) of PMMA and PMMA composites were also analysed by TMA and values obtained is in agreement with the values obtained by DSC. The Table 1 gives the values obtained for the pure PMMA and the composites

*3.3. Differential Scanning Calorimetry (DSC):*

Differential Scanning Calorimetry (DSC) measurements were carried out to determine the thermal properties such as melting temperature ($T_m$), heat of crystallisation ($H_c$) and glass transition ($T_g$) of PMMA and PMMA composites. The $T_g$ of PMMA would vary depending on the degree of tacticity (isotactic, heterotactic, syndiotactic or atactic), thermal history and the type of free radical initiator used during processing [22,23]. The glass transition temperature ($T_g$) of "atactic PMMA" is 105°C, for "isotactic PMMA" it lies between 38°C and 57°C, between 105°C and 138°C for "syndiotactic PMMA" and for the commercial grade PMMA, it ranges from 85 to 165 °C, which are copolymers with co-monomers other than methyl methacrylate [23]. The PMMA used in our work is mostly of "syndiotactic PMMA" type as the commercially available PMMA contains more than one type, in which "syndiotactic" type is predominant. Fig.6 shows the differential scanning calorimetry (DSC) traces obtained for PMMA and PMMA+CCTO composites. The endothermic peak corresponding to PMMA could not be detected (Fig.6); it has however exhibited a heat flow change at approximately 107°C, corresponding to the glass transition ($T_g$) of "syndiotactic PMMA". The DSC study indicated that there is no significant change in the heat flow behaviour between the pure PMMA and the composites. The glass transition ($T_g$) temperature either increases or decreases depending on the type of filler, its morphology and the intercrystallites distance [24-26]. However, we did not observe any change in the glass transition ($T_g$) upto 38 vol % loading of CCTO in PMMA.

*3.4. FTIR studies:*

FTIR analysis on PMMA has been well documented [27-29]. Also, the recent investigations into the vibrational spectra and normal coordinate analysis yielded a complete and very clear information on the fundamental vibrations of PMMA structure [28]. The FTIR spectra recorded for the pure PMMA and for the present composite samples are depicted in Figs.7 and 8 respectively. For PMMA (fig.7), the absorption peaks around 2992 cm$^{-1}$ and 2949 cm$^{-1}$ correspond to C-H asymmetric stretching in $CH_3$ and $CH_2$ respectively. The vibrational band at 2847 cm$^{-1}$ is due to the C-H symmetric stretching in $CH_3$. The characteristic band for the pure PMMA is observed at 1721 cm$^{-1}$, which corresponds to C=O stretching band. The vibrations due to deformation modes of $CH_3$ groups appear at 1489 cm$^{-1}$, at 1434 cm$^{-1}$ and at 1385 cm$^{-1}$. Medium bands at 1268 cm$^{-1}$ and at 1238 cm$^{-1}$ correspond to C-O stretching modes. The band at 1189 cm$^{-1}$ corresponds to $CH_3$ wagging and two bands at 1141 cm$^{-1}$ and 1062 cm$^{-1}$ are due to the $CH_3$ twisting respectively. The vibration modes due to C-C stretching appear at 985 cm$^{-1}$ and 969 cm$^{-1}$. The peaks at 912 cm$^{-1}$ and 840cm$^{-1}$ are assigned to $CH_2$ rocking and the peaks at 809 and 749 are due to the C=O in plane and out of plan bending respectively [28]. The FTIR spectra of pure PMMA(fig.7) and the composite (fig.8) did not show any appreciable change in the 600-4000 cm$^{-1}$ region. However, there is an appreciable change in the absorption in the 400 to 600 cm$^{-1}$ region which is possibly owing to the interaction of CCTO ceramic powder with the PMMA matrix. There is a distinctive change of characteristic absorption in the region 400 to 600 cm$^{-1}$ (fig.8), which are due to the addition of CCTO. There are some new multiple absorption bands persisting in the region 600-400 cm$^{-1}$ arising mainly from the M-O (M=Metal, O=Oxygen) bands in CCTO [9] and the relative intensity of peaks in this region (400 to 600 cm$^{-1}$) has increased, indicating that there is a good interaction between the CCTO and PMMA matrix.

*3.5 Frequency dependent room temperature permittivity*

The room temperature effective permittivity data ($\varepsilon_{eff}$) for PMMA/CCTO composites for different volume percents of CCTO are given in figure.9. The permittivity of the pure PMMA is around 4.9 @100Hz, which is nearly constant over the entire frequency range (100Hz-100MHz) covered in the present investigation (fig.9). As expected, the permittivity increases as the ceramic loading increases in the polymer matrix at all the frequencies under study. The permittivity has increased to 15.7@100Hz when the ceramic loading is increased to 38 Vol % in PMMA. The value of permittivity obtained for the PMMA+38 Vol % CCTO is higher than that of the pure PMMA and much lower than that of CCTO ceramic (fig.9). The room temperature dielectric loss (D) recorded as a function of frequency is shown in fig.10. The dielectric loss increased as the CCTO content increased in PMMA, but decreased with the increase in frequency. Around 10 kHz, there is a sudden drop in the value of loss. It is known that the addition of fillers induces structural changes in the PMMA matrix [30] which may result in a sudden drop in the dielectric loss in the present composites. Indeed this is corroborated by the additional peaks that are encountered in the present FTIR studies. The higher dielectric loss especially at low frequencies, is attributed to interfacial polarization/MWS effect [17,30]. The dielectric loss for all the composites lies below 0.1 for the entire frequency range under investigation. The dielectric loss obtained for pure PMMA is almost independent of the frequency (100Hz to 100MHz). In this work, it is to be noted that the dielectric loss obtained in CCTO/PMMA composites is remarkably low. For instance, the loss value obtained @100 Hz for PMMA+38 Vol % CCTO composite is around 0.094 and has decreased to 0.011 @100MHz, demonstrating that this PMMA+CCTO composite may be exploited in the design and fabrication of capacitors for high frequency applications.

The AC conductivity $\sigma$ is derived from the dielectric data using the equation (7):

$$\sigma' = \varepsilon_o \omega \varepsilon'' \tag{7}$$

Where $\varepsilon_o$ = 8.85.10$^{-12}$ F/m is the permittivity of the free space and $\omega = 2\pi f$ the angular frequency. The variation of AC conductivity as a function of frequency for the PMMA and the composites is shown in Fig.11. It has been observed that the AC conductivity increases with the frequency, and as reported [30], the increase in AC conductivity at low frequency is due to the interfacial polarization while in the high frequency region, it is attributed to the electronic polarization [35].

## CONCLUSIONS

The PMMA-CCTO composite sheets have been successfully fabricated via melt mixing and hot pressing technique. In order to obtain a flexible composite, the ceramic loading has been restricted to a maximum of 38 Vol %. The composites exhibited better thermal stability than that of the pure PMMA, though there is no change in the glass transition ($T_g$) between the polymer and the composites. The permittivity of PMMA increased with increase in CCTO content associated with low dielectric loss at high frequencies. The present composite sheets of PMMA+CCTO may be exploited in the design and fabrication of capacitors for high frequency applications.

## ACKNOWLEDGEMENT

The management of Central Power Research Institute is acknowledged for the financial support (CPRI Project No. R-DMD-01/1415).

## FIGURE CAPTIONS

Figure.1 Flowchart depicting various steps involved in the fabrication of Composite-A.

Figure.2 X-ray diffraction patterns for (a) pure PMMA, (b) $1000^{o}C/10h$ –Phase-pure CCTO, (c) CCTO-JCPDS, (d) PMMA+6 Vol % CCTO and (e) PMMA+38 Vol % CCTO.

Figure.3. Thermal analysis (TG) for the (a) pure PMMA, (b) 6, (c) 10, (d) 21, (e) 28 and (f) for 38 Vol % PMMA-CCTO composite.

Figure.4. Derivative Thermo Gravimetric (DTG) analysis for the (a) pure PMMA, (b) 6, (c) 10, (d) 21, (e) 28 and (f) for 38 Vol % PMMA-CCTO composite.

Figure.5. DSC traces for the (a) pure PMMA, (b) 6, (c) 10, (d) 21, (e) 28 and (f) for 38 Vol % PMMA-CCTO composite.

Figure.6. Thermo Mechanical Analysis (TMA) for the (a) pure PMMA, (b) 6, (c) 10, (d) 21, (e) 28 and (f) for 38 Vol % PMMA-CCTO composite.

Figure.7. FTIR spectra recorded for the pure PMMA.

Figure.8. FTIR spectra recorded for (a) 6, (b) 10, (c) 21, (d) 28 and (e) for 38 Vol % PMMA-CCTO composite.

Figure.9. Frequency dependent of dielectric permittivity measured at room temperature (300K) for different Vol % PMMA-CCTO composite.

Figure.10 Frequency dependent dielectric loss measured at room temperature (300K) for different Vol % PMMA-CCTO composite

Figure.11. Frequency dependent AC conductivity for different volume percents of CCTO.

*Table.1. Density, CTE, glass transition temperature, dielectric permittivity and loss values obtained for the pure PMMA and the composites.*

| Sample | Density, g/cm$^3$ | Coefficient of Thermal Expansion (CTE), (μm / m.°C), | Glass Transition($T_g$)/ Temperature, °C | Permittivity @ 10kHz | Dielectric loss @ 10kHz |
|---|---|---|---|---|---|
| PMMA | 1.180 | 93.10 | 107.94 | 4.05 | 0.016 |
| PMMA+CCTO-6 | 1.421 | 87.99 | 107.72 | 4.20 | 0.035 |
| PMMA+CCTO-10 | 1.576 | 53.14 | 107.80 | 4.51 | 0.037 |
| PMMA+CCTO -21 | 1.722 | 53.86 | 107.85 | 9.13 | 0.040 |
| PMMA+CCTO -28 | 1.894 | 51.06 | 107.86 | 11.61 | 0.042 |
| PMMA+CCTO -38 | 2.055 | 48.25 | 106.33 | 12.82 | 0.045 |

Figure.1
Click here to download high resolution image

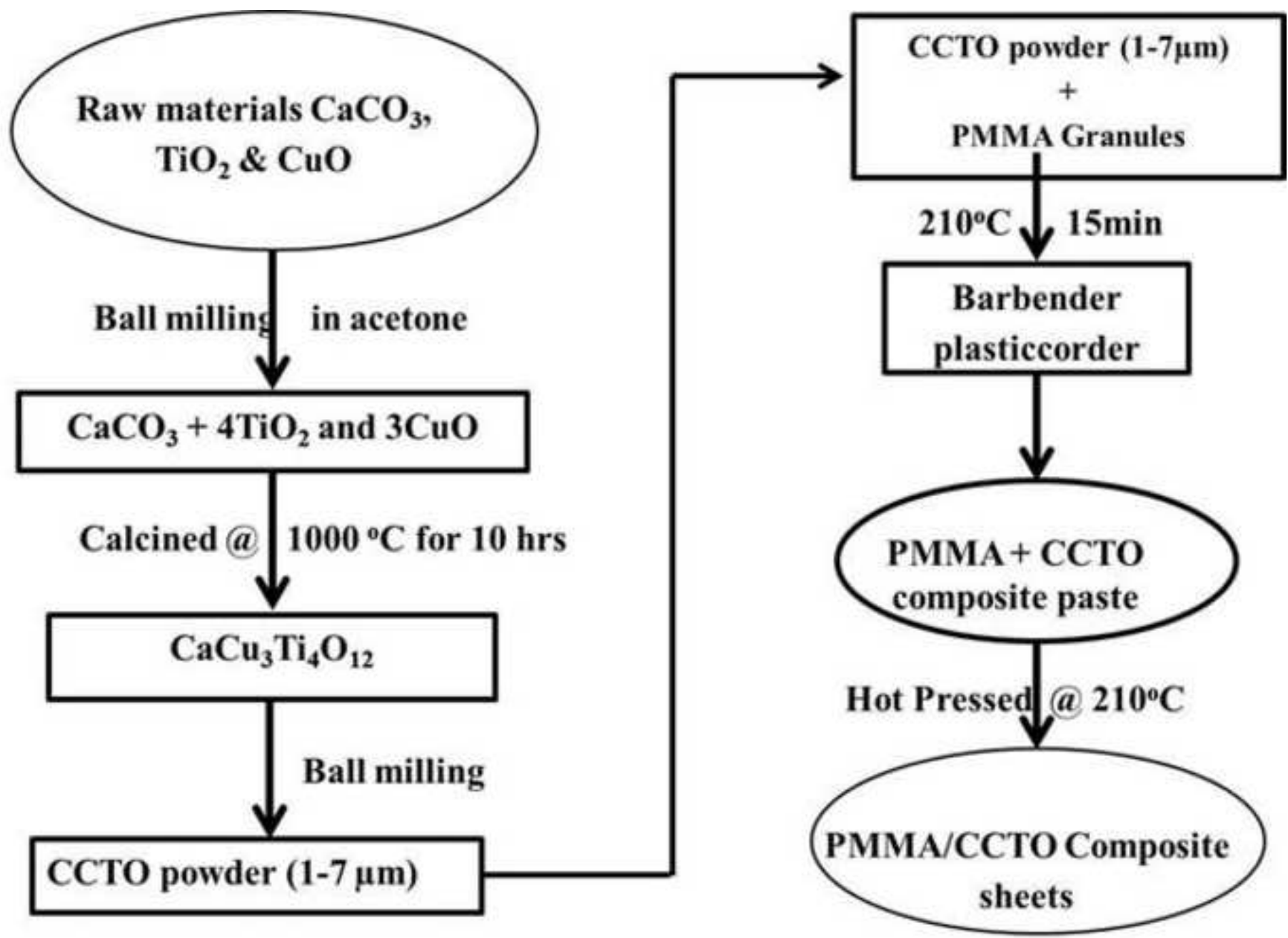

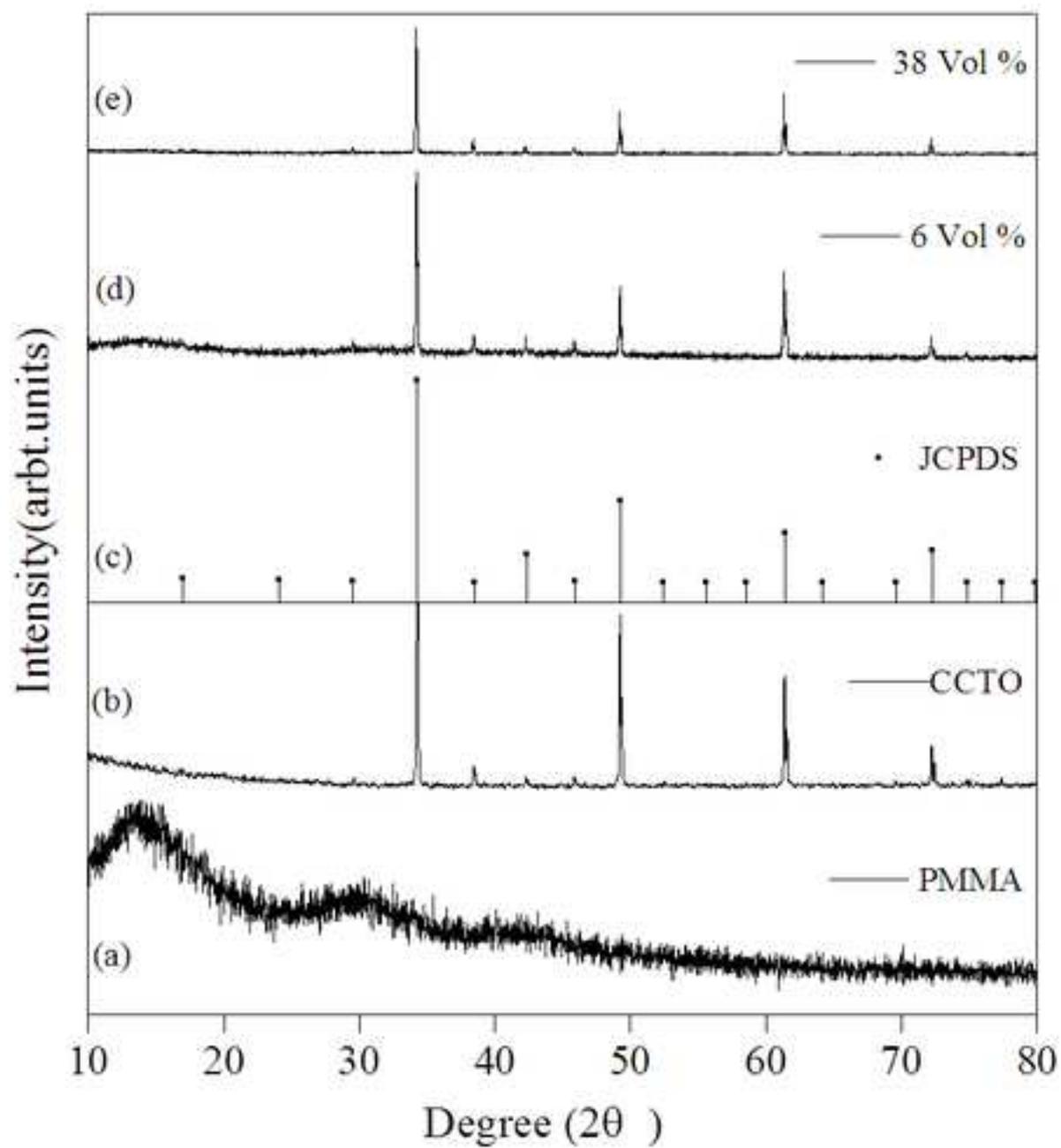

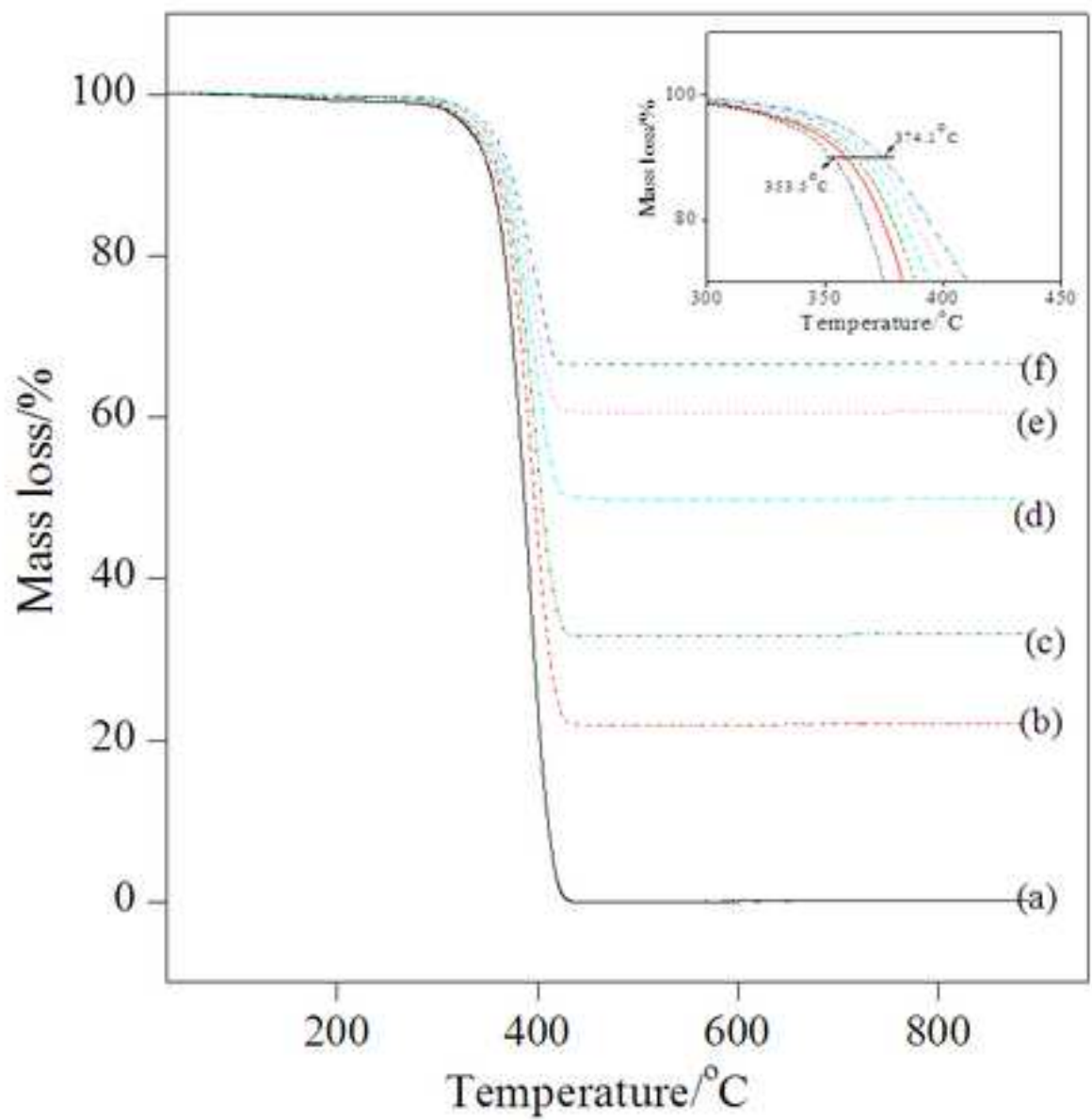

Figure.3

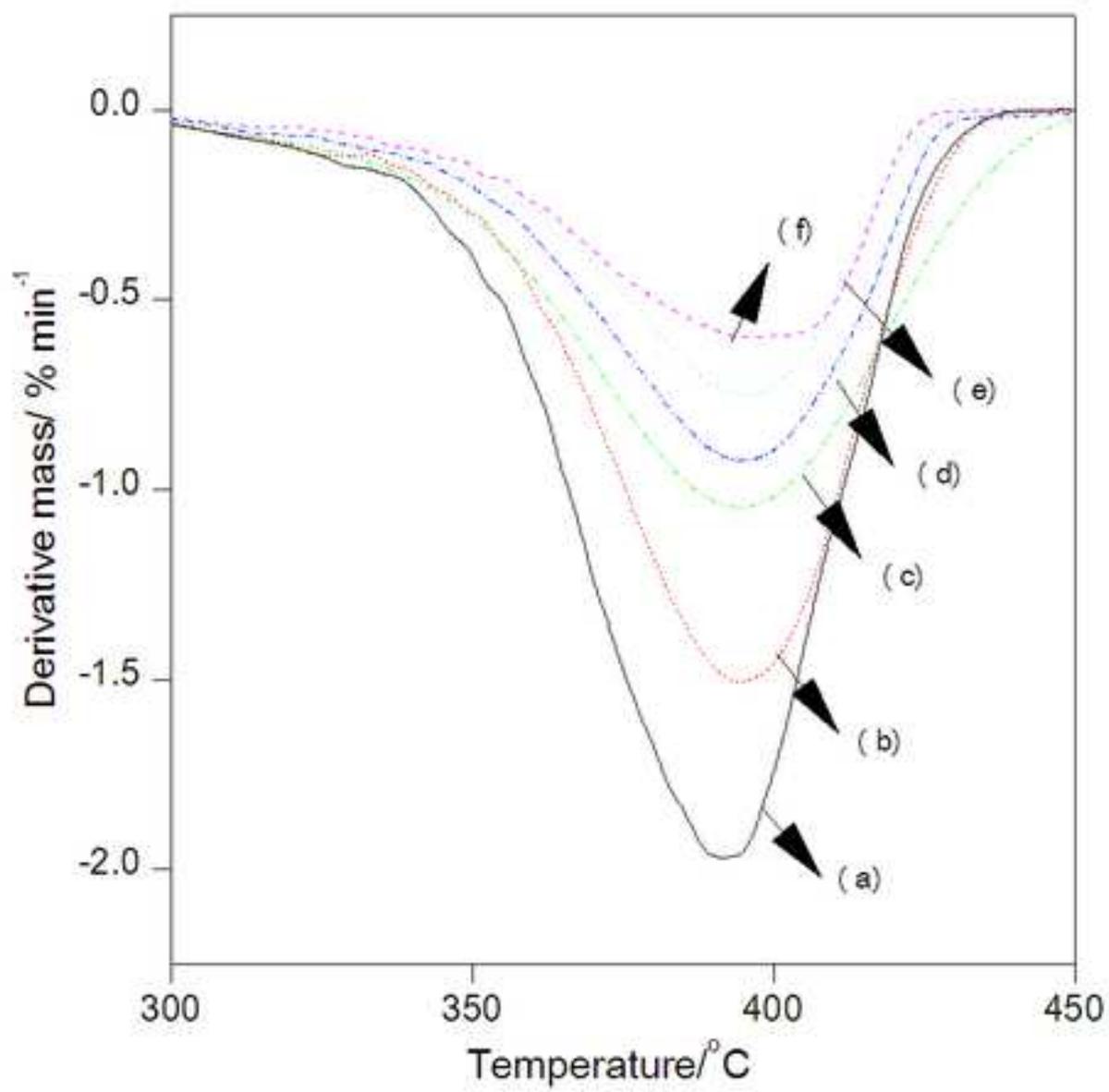

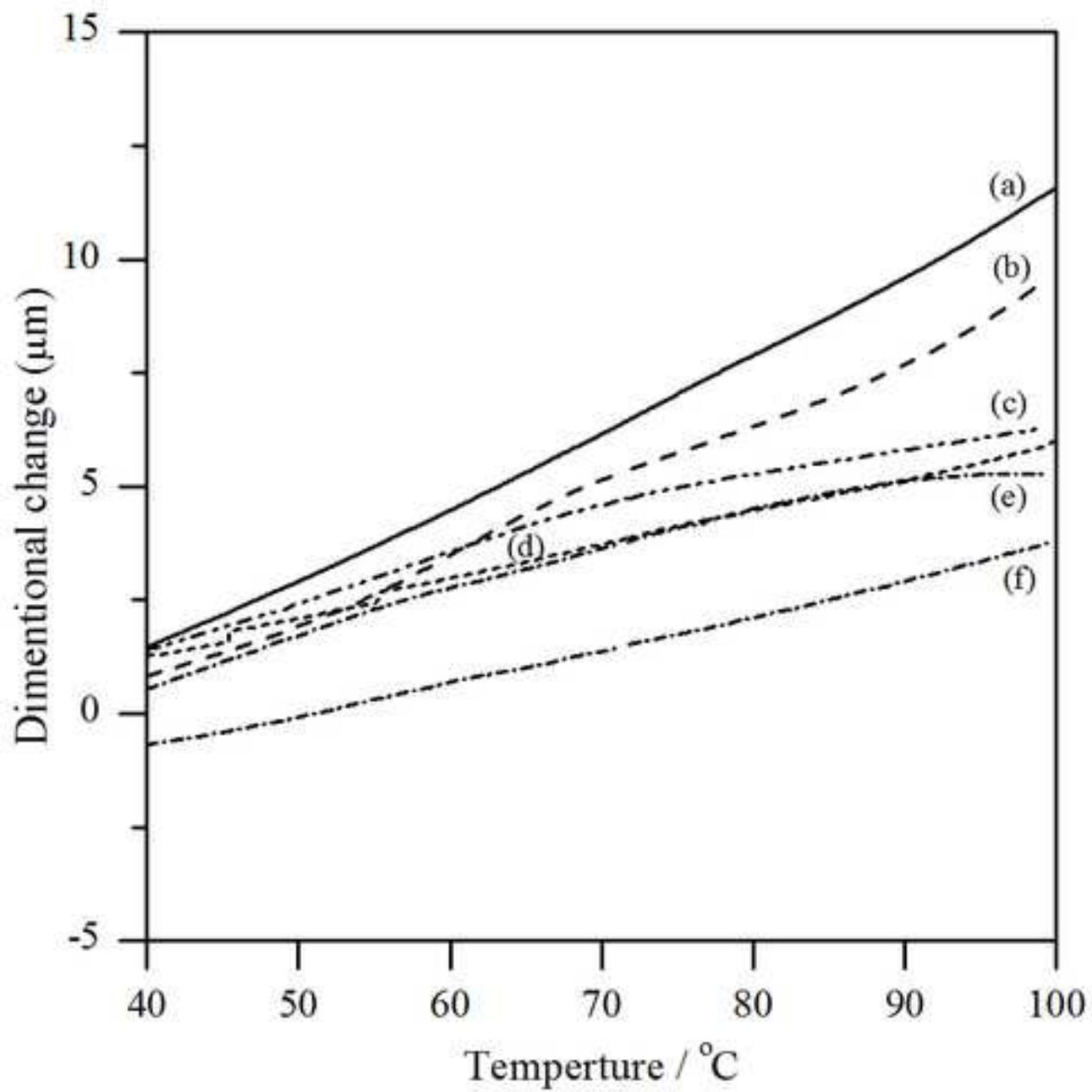

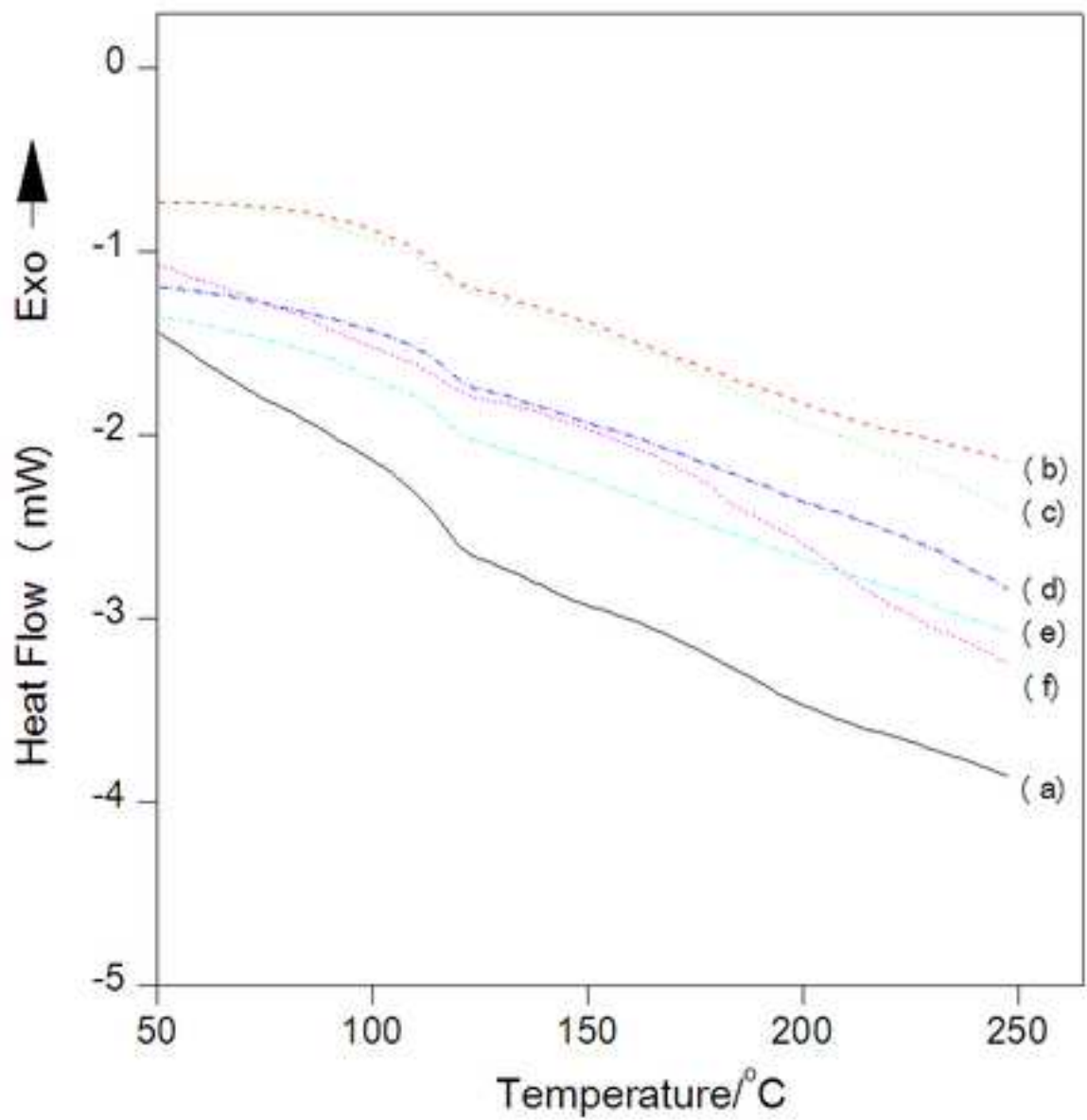

Figure.7
Click here to download high resolution image

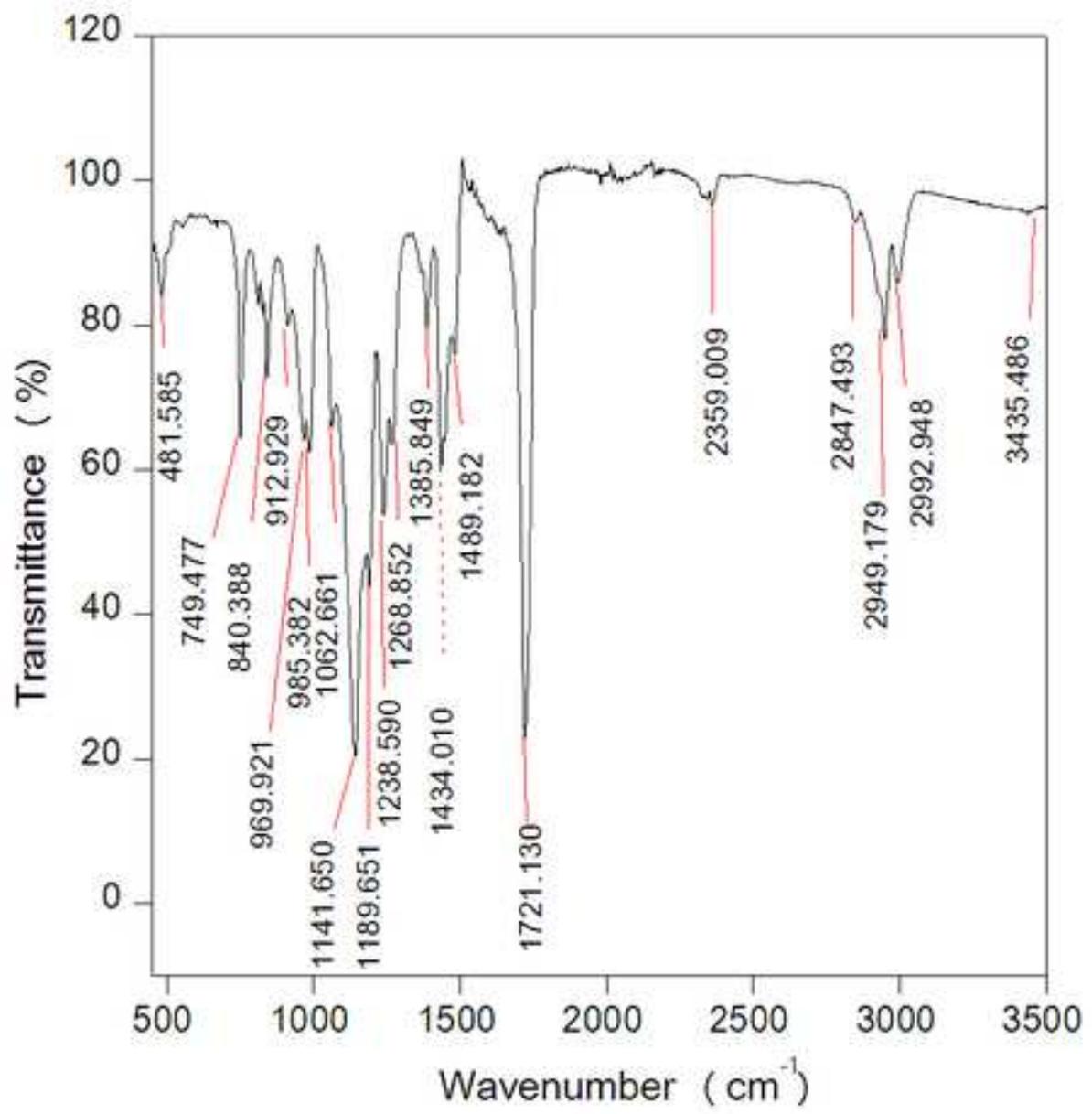

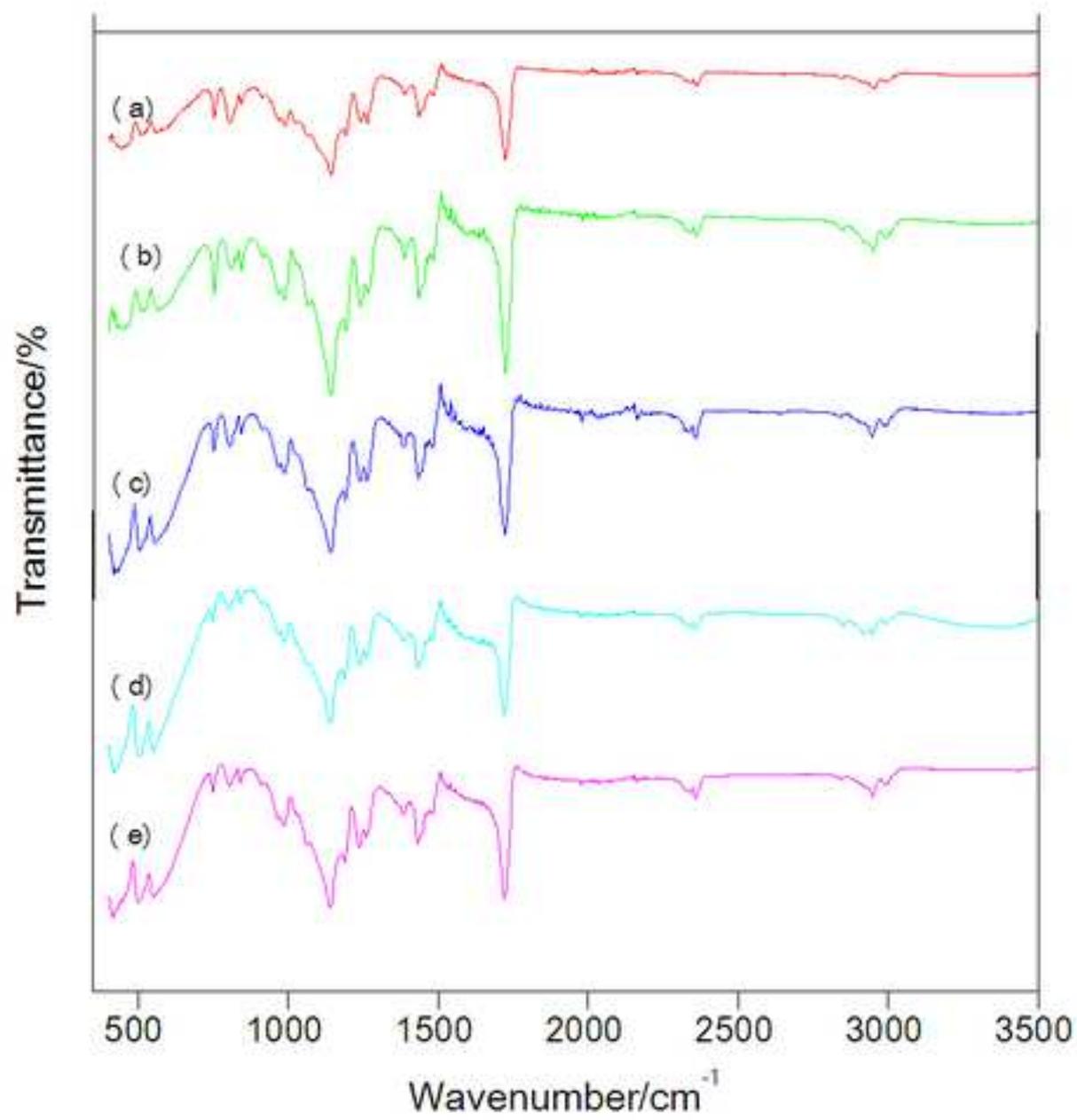

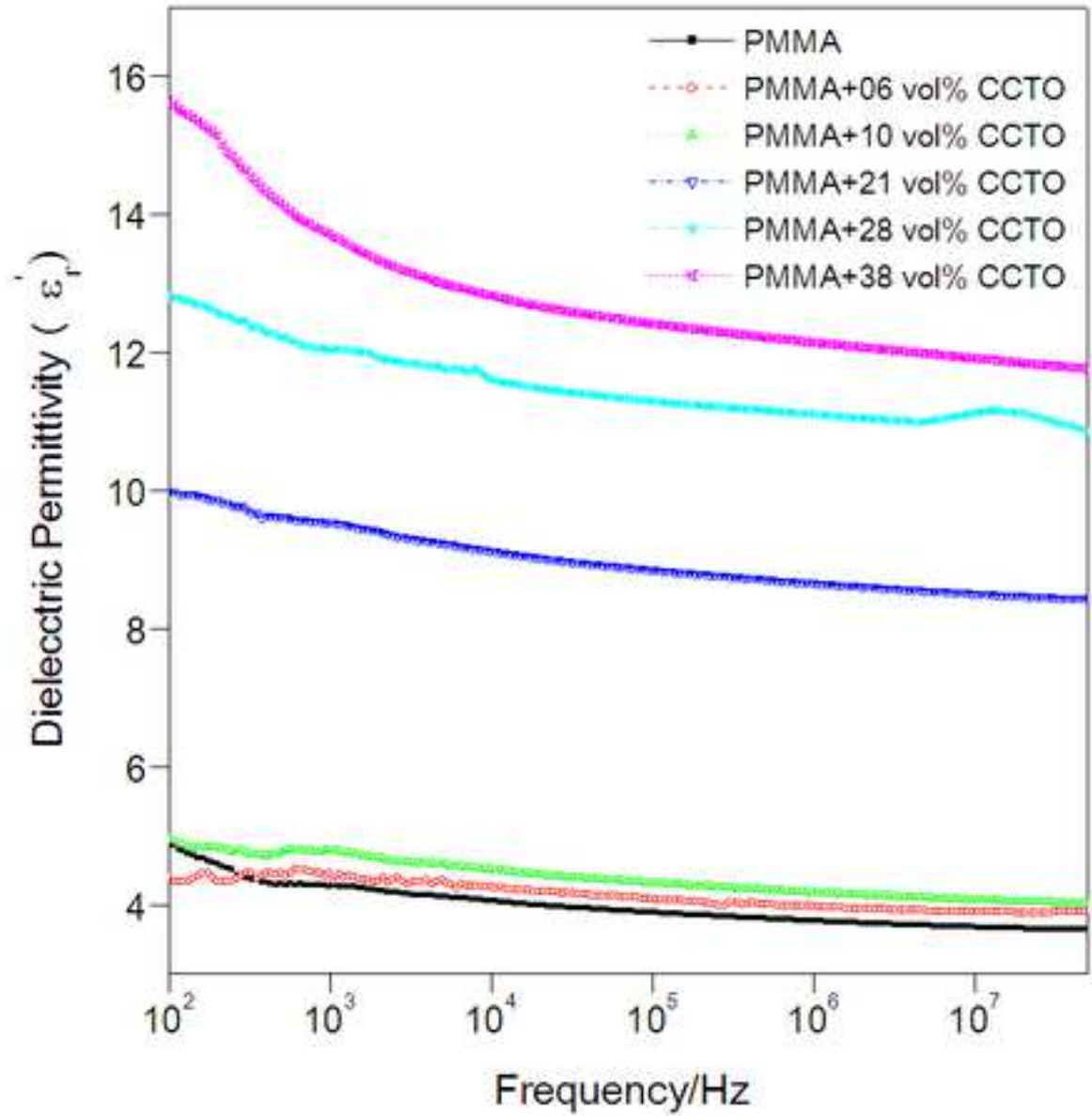

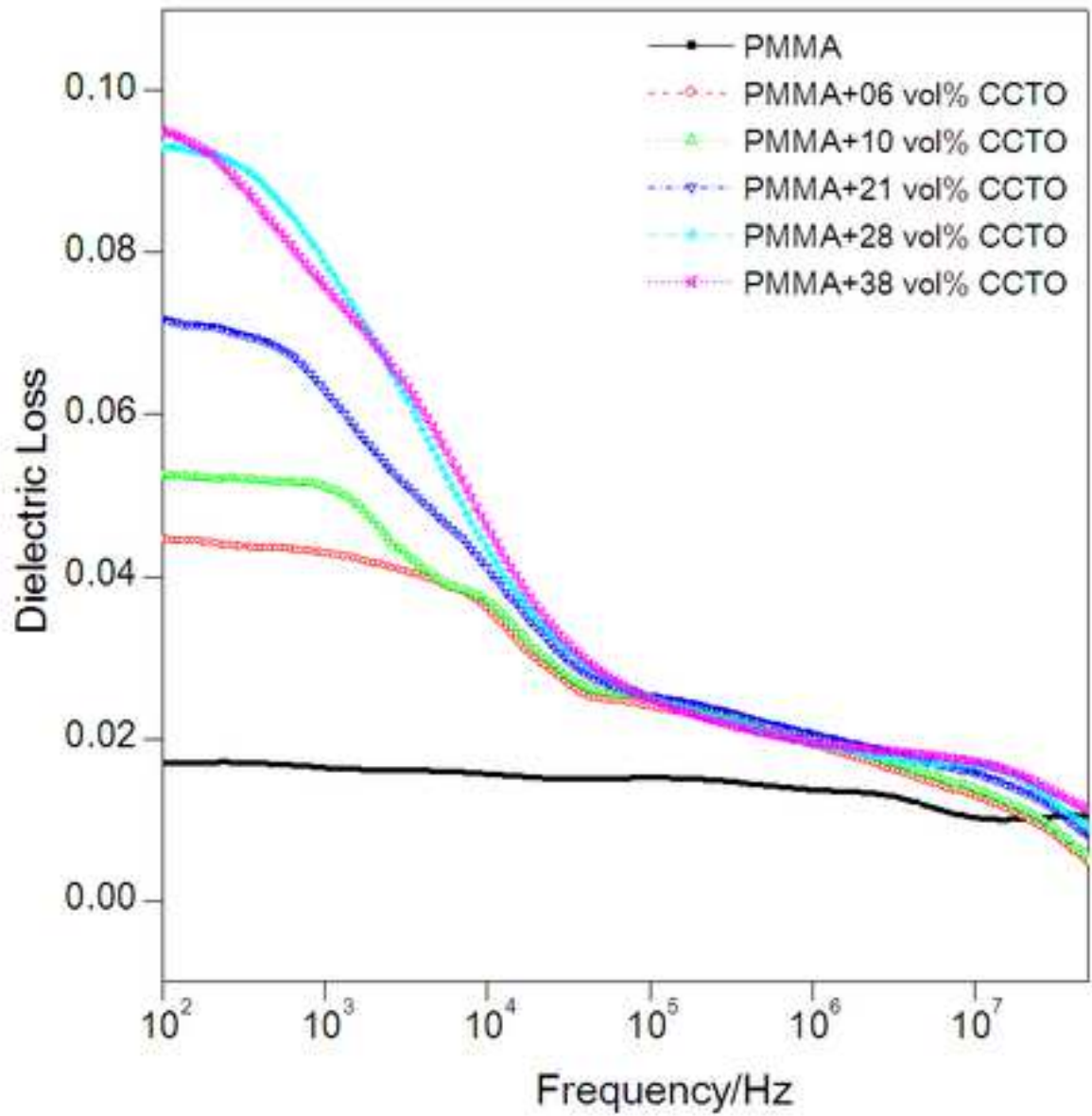

Figure.11
Click here to download high resolution image

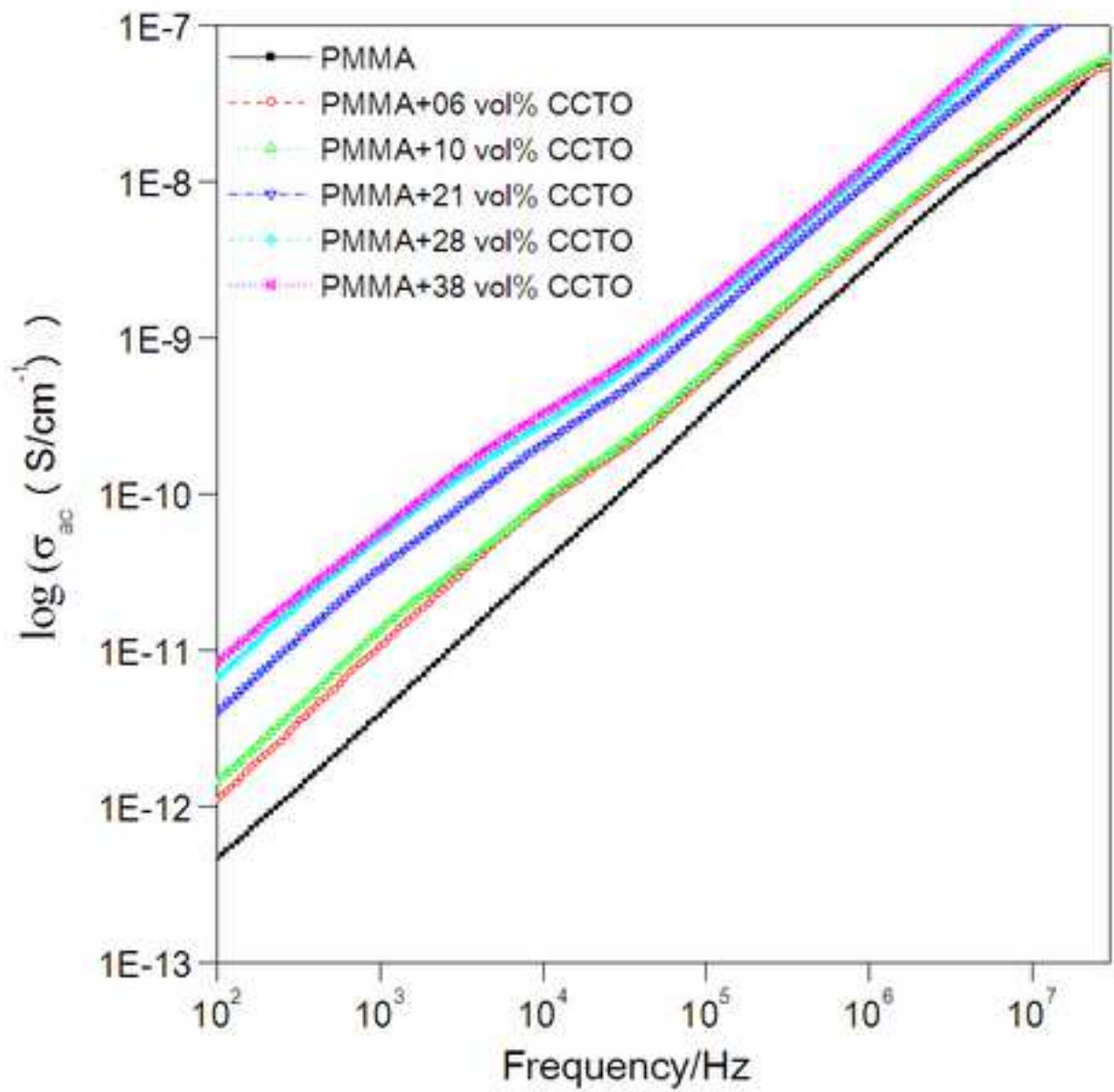